\documentclass[twocolumn,showpacs,amsmath,aps,prl]{revtex4}

 \usepackage{graphicx}
 \usepackage{bm}

\begin{document}
\title{Recursion relations for generalized Fresnel coefficients:
Casimir force in a planar cavity}
 \author{Marin-Slobodan Toma\v s}
 \email{tomas@thphys.irb.hr}
 \affiliation{Rudjer Bo\v skovi\' c Institute, P. O. B. 180,
 10002 Zagreb, Croatia}
 \date{\today}

 \begin{abstract}
We emphasize and demonstrate that, besides using the usual
recursion relations involving successive layers, generalized
Fresnel coefficients of a multilayer can equivalently be
calculated using the recursion relations involving stacks of
layers, as introduced some time ago [M. S. Toma\v s, Phys. Rev. A
{\bf 51}, 2545 (1995)]. Moreover, since the definition of the
generalized Fresnel coefficients employed does not imply
properties of the stacks, these nonstandard recursion relations
can be used to calculate Fresnel coefficients not only for a local
but also for a general multilayer consisting of various types
(local, nonlocal, inhomogeneous, etc.) of layers. Their utility is
illustrated by deriving a few simple algorithms for calculating
the reflectivity of a Bragg mirror and extending the formula for
the Casimir force in a planar cavity to arbitrary media.

\end{abstract}
 %\preprint{IRB-TH-4/06}
 \maketitle

Generalized Fresnel coefficients are basic ingredients in the
theory of electromagnetic processes and effects in layered systems
such as light propagation in stratified media \cite{BW,Chew},
molecular fluorescence and energy-transfer near interfaces
\cite{CPS}, (dipole) radiation from multilayers
\cite{Luk,Reed,Craw}, spontaneous emission and light scattering at
surfaces \cite{FW} and in planar cavities\cite{DeM,Tom95}, the
Casimir effect between multilayered stacks \cite{Zhou,Tom02,Pars}
etc. Correspondingly, the problem of calculating Fresnel
coefficients arises in many area of physics: optics, surface
physics, spectroscopy of multilayers, cavity QED, theory of the
Casimir effect etc. For systems consisting of few layers, these
coefficients are standardly calculated using (ordinary) recursion
relations involving successive layers. With increasing number of
layers, however, this method soon leads to cumbersome formulas and
thus becomes impractical. Therefore, despite the possibility of a
polynomial representation of the generalized Fresnel coefficients
\cite{Vig}, for more complex systems these coefficients are
conveniently calculated using the transfer matrix method
\cite{BW,Chew,Reed,Pars}.

\begin{figure}[htb]
 \begin{center}
 \resizebox{6cm}{!}{\includegraphics{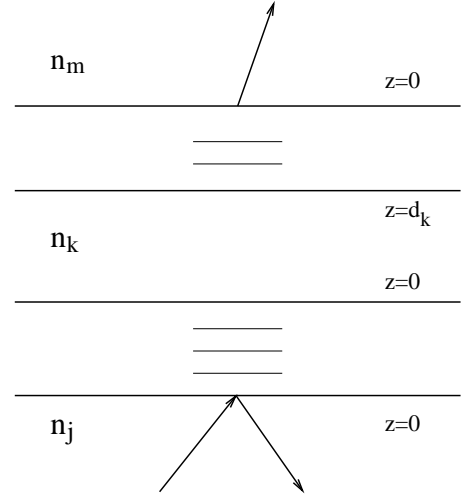}}
 \end{center}
 \caption{\label{sys}Stack considered when deriving recursion relations for
generalized Fresnel coefficients shown schematically. Layers $j$,
$k$ and $m$ are local and described by (complex) refraction
indexes $n_a(\omega)=\sqrt{\varepsilon_a(\omega)\mu_a(\omega)}$,
$a=j,k,m$, whereas stacks between them are unspecified. Arrows
indicate propagation of a wave incident on the stack. A
shifted-$z$ coordinate system is adopted, as explained in the
text.}
\end{figure}

Based on the definition of the generalized Fresnel coefficients
given in Ref. \cite{Craw}, in our consideration of the Green
function for a (local) multilayered system \cite{Tom95} we have
used the recursion relations for the reflection ($r$) and
transmission ($t$) coefficients involving {\it stacks} of layers,
which enabled us to write the Green function in a simple compact
form. For a stack of layers between, say, layers $j$ and $m$
(denoted shortly as $j/m\equiv j\ldots m$) these recursion
relations read \cite{Tom95} (unless necessary, we omit the
polarization index ${q=p,s}$)
\begin{equation}
r_{j/m}\equiv r_{j/k/m}=r_{j/k}+\frac{t_{j/k}t_{k/j}r_{k/m}
 e^{2i\beta_kd_k}}{1-r_{k/j}r_{k/m}e^{2i\beta_kd_k}},
 \label{rj-k-m}
 \end{equation}
\begin{equation}
t_{j/m}\equiv t_{j/k/m}=\frac{t_{j/k}t_{k/m}e^{i\beta_kd_k}}
{1-r_{k/j}r_{k/m}e^{2i\beta_kd_k}}.
\label{tj-k-m}
\end{equation}
where $k$ is an intermediate layer, as depicted in Fig. \ref{sys},
and where by using the notation $r_{j/k/m}$ we simply stress to
which intermediate layer we address. As seen, the above recurrence
relations look the same as the standard ones \cite{BW,Chew} (to
which they reduce in case of a system $jk/m\equiv jk\ldots m$),
however, this time they generally involve Fresnel coefficients for
stacks of layers.

Although Eqs. (\ref{rj-k-m}) and (\ref{tj-k-m}) have also been
known for some time in optics of multilayers \cite{Kat}, it seems
that the possibility of grouping the layers in stacks when
calculating Fresnel coefficients is not widely recognized. Since
this method is particularly convenient in some cases, e.g., when
calculating Fresnel coefficients of periodic media, in this work
we explicitly derive the above recurrence relations starting from
the definition of the generalized Fresnel coefficients \cite{Craw}
(this derivation was omitted in Ref. \cite{Tom95}) and demonstrate
their equivalence with the standard recursion relations involving
successive layers. We use this opportunity to emphasize that Eqs.
(\ref{rj-k-m}) and (\ref{tj-k-m}) are also valid for piecewise
nonlocal stratified media and adapt their derivation accordingly.
We illustrate their usefulness on a few simple examples including
the theory of the Casimir effect in (nonlocal) multilayers.

Consider a linear system consisting of isotropic layers $ 0\ldots
n\equiv 0/n$. The electric field of a wave incident on the system
with the parallel wave vector ${\bf k}$ can in a local layer $l$
be written as
 \begin{eqnarray}
{\bf E}_l({\bf k},\omega,{\bf r})&=&{\bf E}_l({\bf
k},\omega,z)\exp(i{\bf k}\cdot {\bf r}_\parallel),\nonumber\\
 {\bf E}_l({\bf k},\omega,z)&=&
\hat{\bf e}^+_l({\bf k})e^{i\beta_lz}E^+_l
 +\hat{\bf e}^-_l({\bf k})e^{-i\beta_lz}E^-_l,
 \label{El}
 \end{eqnarray}
where $\beta_l=\sqrt{k^2_l-k^2}$, with
$k_l=n_l\omega/c=\sqrt{\varepsilon_l\mu_l}\omega/c$. For a local
system, the coefficients $E^\pm_l$ are determined by matching the
field at boundaries of each layer and can be entirely expressed in
terms of the incident-wave amplitude and the generalized
reflection and transmission coefficients of the corresponding
stacks of layers \cite{Craw,Tom95}.

We define generalized Fresnel coefficients of a stack between two
(local) layers as follows \cite{Craw}. A reflection coefficient
$r$ of a stack is the ratio of the reflected to incoming wave
electric-field amplitude (factors multiplying $\hat{\bf e}$'s) at
the corresponding stack's boundary considering only layers of the
stack, i.e. as if there are no other layers in the system.
Similarly, a transmission coefficient $t$ of a stack is the ratio
of the transmitted to incident wave amplitude calculated at the
corresponding stack's boundaries as if there are no other layers
present in the system. Note that these definitions do not imply
any property of the stack. Therefore, we can use these
coefficients to describe wave propagation not only in a local
system but also in a general system consisting of various (local,
nonlocal, inhomogeneous, unspecified etc.) layers. In calculating
these coefficients it is convenient to adopt a (shifted-$z$)
representation for the field \cite{Tom95} in which $0\leq z\leq
d_l$ in any finite layer $l$, whereas $-\infty<z\leq 0$  and
$0\leq z<\infty$ in external layers $l=0$ and $l=n$, respectively.

On the basis of the above definition we can straightforwardly
calculate the recurrence relations for generalized Fresnel
coefficients $r_{j/m}$ and $t_{j/m}$ of the stack of layers
between local layers $j$ and $m$. Indeed, according to its
definition, to calculate the reflection coefficient $r_{j/m}$ we
consider the wave incident from the layer $j$ upon the system
$j/m$ (see Fig. \ref{sys}). Then $E_m^-=0$ and by definition
\begin{equation}
\label{rjm}
 E^-_j=r_{j/m}E^+_j.
 \end{equation}
However, $r_{j/m}$ can also be calculated by considering
transmission of the wave to an intermediate (local) layer $k$ and
its subsequent partial transmission back to the layer $j$.
Considering the field at the relevant boundaries and using the
above definitions of the generalized Fresnel coefficients, the
amplitudes of the wave are then related by the following set of
equations
\begin{subequations}
\label{BC}
\begin{equation}
\label{BCa}
 E^-_j=r_{j/k}E^+_j+t_{k/j}E^-_k,
\end{equation}
\begin{equation}
 E^+_k=t_{j/k}E^+_j+r_{k/j}E^-_k,
\end{equation}
\begin{equation}
e^{-i\beta_kd_k}E^-_k=r_{k/m}e^{i\beta_kd_k}E^+_k.
 \end{equation}
 \end{subequations}
Eliminating $E^\pm_k$ and comparing the ratio $E^-_j/E^+_j$ with
that from Eq. (\ref{rjm}), we arrive at Eq. (\ref{rj-k-m}).
Similarly, the transmission coefficient $t_{j/m}$ of the stack is
by
 definition given by
\begin{equation}
\label{tjl}
 E^+_m=t_{j/m}E^+_j.
 \end{equation}
On the other hand, by considering transmission of the wave to the
intermediate layer $k$ and its subsequent partial transmission to
the layer $m$, we find that the amplitudes $E^+_m$ and $E^+_j$ are
related by Eq. (\ref{BC}), with Eq. (\ref{BCa}) replaced by
\begin{equation}
E^+_m=t_{k/m}e^{i\beta_kd_k}E^+_k.
 \end{equation}
Proceeding as before, we obtain Eq. (\ref{tj-k-m}).

Clearly, Fresnel coefficients must not depend on the choice of the
intermediate layer in Eqs. (\ref{rj-k-m}) and (\ref{tj-k-m}).
Therefore, to prove the consistency of these recurrence relations,
we must show that
\begin{equation}
\label{rt} r_{j/k/m}=r_{j/l/m},\;\;\;t_{j/k/m}=t_{j/l/m},
\end{equation}
where $l$ denotes some other intermediate local layer. To prove
this for $r_{j/m}$, we rewrite Eq. (\ref{rj-k-m}) in the form
\begin{eqnarray}
 r_{j/m}&=&r_{j/k/m}=\frac{r_{j/k}+a_{j/k}
 r_{k/m}e^{2i\beta_{k}d_{k}}}
 {1-r_{k/j}r_{k/m}e^{2i\beta_{k}d_{k}}},\nonumber\\
 a_{j/k}&=&t_{j/k}t_{k/j}-r_{j/k}r_{k/j}=a_{k/j},
 \label{rj-k-m2}
 \end{eqnarray}
consider the layer $l$ between layers $k$ and $m$ and apply Eq.
(\ref{rj-k-m2}) to the reflection coefficient $r_{k/m}=r_{k/l/m}$
in this very same equation. Rearranging the terms and using again
Eq. (\ref{rj-k-m2}) to recognize reflection coefficients $r_{j/l}$
and $r_{l/j}$, we find
\begin{eqnarray}
 r_{j/k/m}&=&\frac{r_{j/l}+\tilde{a}_{j/l} r_{l/m}e^{2i\beta_ld_l}}
 {1-r_{l/j}r_{l/m}e^{2i\beta_ld_l}},\nonumber\\
 \tilde{a}_{j/l}&=&\frac{a_{j/k}a_{l/k}e^{2i\beta_kd_k}-r_{j/k}r_{l/k}}
 {1-r_{k/j}r_{k/l}e^{2i\beta_kd_k}}
 \label{rA}
 \end{eqnarray}
Now, noting from Eqs. (\ref{tj-k-m}) and (\ref{rj-k-m2}) that
 \begin{equation}
 t_{j/l}t_{l/j}=\frac{(a_{j/k}+r_{j/k}r_{k/j})(a_{l/k}+r_{l/k}r_{k/l})
 e^{2i\beta_kd_k}}{(1-r_{k/j}r_{k/l}e^{2i\beta_kd_k})^2}
 \end{equation}
and that $r_{j/l}r_{l/j}$ is given by a similar expression, we
find that actually \cite{Vig2}
\begin{equation}
\label{ata} a_{j/l}=t_{j/l}t_{l/j}-r_{j/l}r_{l/j}=\tilde{a}_{j/l}.
\end{equation}
Accordingly, the right-hand side of Eq. (\ref{rA}) is, upon using
Eq. (\ref{rj-k-m2}), indeed identified as $r_{j/l/m}$. To prove
the equality of the transmission coefficients in Eq. (\ref{rt}) we
first apply Eq. (\ref{tj-k-m}) to the coefficient
$t_{k/m}=t_{k/l/m}$ and then use the identity
($r_{k/m}=r_{k/l/m}$) \cite{Tom02}
\begin{eqnarray}
\label{DD}
&(1-r_{l/k}r_{l/m}e^{2i\beta_ld_l})(1-r_{k/j}r_{k/m}e^{2i\beta_kd_k})=\nonumber\\
&(1-r_{k/j}r_{k/l}e^{2i\beta_kd_k})(1-r_{l/j}r_{l/m}e^{2i\beta_ld_l}),
\end{eqnarray}
which follows from Eq. (\ref{rj-k-m2}). In this way, we obtain
\begin{equation}
 t_{j/k/m}=\frac{t_{j/k}t_{k/l}t_{l/m}
 e^{i(\beta_kd_k+\beta_ld_l})}{(1-r_{k/j}r_{k/l}e^{2i\beta_kd_k})
 (1-r_{l/j}r_{l/m}e^{2i\beta_ld_l})},
 \end{equation}
which is according to Eq. (\ref{tj-k-m}) equal to $t_{j/l/m}$.

In most textbook approaches to the wave propagation in layered
media the basic ingredients in the calculation of the generalized
Fresnel coefficients are coefficients $r_{jk}$ and $t_{jk}$ for
the interface between two neighbouring local media $j$ and $k$.
With unit polarization vectors in Eq. (\ref{El}) \cite{Tom95}
\begin{equation}
\hat{\bf e}^\pm_{pl}=\frac{\mp \beta_l\hat{\bf k}+k\hat{\bf
z}}{k_l},\;\;\;\hat{\bf e}^\pm_{sl}=\hat{\bf k}\times\hat{\bf z},
\label{eps}
\end{equation}
and applying the usual boundary conditions for the field, it is
straightforward to show that the above definition of Fresnel
coefficients leads to the standard single-interface coefficients
 \begin{subequations}
 \label{sic}
 \begin{equation}
 \label{rjk}
 r_{jk}=\frac{\beta_j-\gamma_{jk}\beta_k}
 {\beta_j+\gamma_{jk}\beta_k}=-r_{kj},
 \end{equation}
 \begin{equation}
 \label{tjk}
 t_{jk}=\sqrt{\frac{\gamma_{jk}}{\gamma^s_{jk}}}(1+r_{jk})=
 \frac{\mu_k\beta_j}{\mu_j\beta_k}t_{kj},
 \end{equation}
 \end{subequations}
 where $\gamma^p_{jk}=\varepsilon_j/\varepsilon_k$ and
 $\gamma^s_{jk}=\mu_j/\mu_k$. As can be easily verified, these
 coefficients obey the Stokes relation
 \begin{equation}
\label{ttrr}
a_{jk}\equiv t_{jk}t_{kj}-r_{jk}r_{kj}=1.
\end{equation}

We note that the symmetry property of the single-interface
transmission coefficient, as expressed by Eq. (\ref{tjk}), implies
for local systems the same symmetry property of the generalized
transmission coefficient $t_{j/m}$, that is \cite{Chew,Tom95}
\begin{equation}
 \label{symtj/m}
 t_{j/m}=\frac{\mu_m\beta_j}{\mu_j\beta_m}t_{m/j}.
 \end{equation}
Indeed, through  Eq. (\ref{tj-k-m}), it certainly holds for a
three-layer system $jkm$ owing to the symmetry property of the
single-interface transmission coefficients $t_{jk}$ and $t_{km}$.
Assuming in Eq. (\ref{tj-k-m}) the same symmetry properties of the
generalized coefficients $t_{j/k}$ and $t_{k/m}$, we immediately
find that Eq. (\ref{symtj/m}) is obeyed. Accordingly, using the
induction argument, we may conclude that this equation is for
local systems generally valid. Now we argue that Eq.
(\ref{symtj/m}) is also valid for piecewise nonlocal systems as it
actually ensures the equality of the stack's transmittances for
waves incident on it from either side. Indeed, the Poynting vector
of the upward/downward-propagating wave in a layer $l$ is
according to Eqs. (\ref{El}) and (\ref{eps}) given by
\begin{equation}
{\bf P}^\pm_l({\bf k},\omega,z)=\frac{c}{8\pi}{\rm
Re}\sqrt{\eta_l}\frac{{\bf k}\pm\beta_l\hat{\bf z}}{k_l}
|E^\pm_le^{\pm i\beta_lz}|^2,
\end{equation}
with $\eta^p_l={\eta^s_l}^*=\varepsilon^*_l/\mu^*_l$. Accordingly,
assuming the outmost layers $j$ and $m$ transparent,
transmittances of the $j/m$ stack for waves incident upward and
downward on it [given by the ratios of the respective transmitted-
to incident-energy fluxes ${\cal T}_{j/m}=P^+_{mz}({\bf
k},\omega,0)/P^+_{jz}({\bf k},\omega,0)$ and ${\cal
T}_{m/j}=P^-_{jz}({\bf k},\omega,0)/P^-_{mz}({\bf k},\omega,0)$]
\begin{equation}
 \label{Tj/m}
 {\cal T}_{j/m}=\frac{\mu_j\beta_m}{\mu_m\beta_j}|t_{j/m}|^2,\hspace{.4cm}
 {\cal T}_{m/j}=\frac{\mu_m\beta_j}{\mu_j\beta_m}|t_{m/j}|^2,
 \end{equation}
can only be equal if Eq. (\ref{symtj/m}) is fulfilled. Thus, being
a consequence of the reciprocity property of the electromagnetic
field, this equation is valid for a quite general class of
(nongyrotropic) media \cite{Pot}.

As follows from the above results, recurrence relations for
Fresnels coefficients of a multilayered system can be written in a
number of ways depending on number of intermediate local layers.
We illustrate this by calculating Fresnel coefficients of few
simple systems. We first consider a $12/3$ system (that is, we
allow for an unspecified stack of layers between layers $2$ and
$3$ of a three-layer local system $123$). Since from Eq.
(\ref{ttrr}) $a_{12}=1$, Eq. (\ref{rj-k-m2}) leads to standard
forms of the recurrence relations for Fresnel coefficients
\cite{Chew}
\begin{equation}
r_{12/3}=\frac{r_{12}+r_{2/3}e^{2i\beta_2d_2}}
 {1-r_{21}r_{2/3}e^{2i\beta_2d_2}},\;\;
 t_{12/3}=\frac{t_{12}t_{2/3}e^{i\beta_2d_2}}
 {1-r_{21}r_{2/3}e^{2i\beta_2d_2}},
\label{r12-3}
 \end{equation}
which, for a three-layer system $123$, reduce to the well-known
results usually quoted in textbooks \cite{BW,Chew}. Next, we
consider Fresnel coefficients of a $123/4$ system. According to
Eq. (\ref{rj-k-m2}), they can be calculated from two equivalent
sets of recurrence relations differing in the choice of the
intermediate layer. The first set is the standard one \cite{Chew}
and is given by Eq. (\ref{r12-3}), with the replacement of indices
$2/3\rightarrow 23/4$. The second set of recurrence relations for
$r_{123/4}$ and $t_{123/4}$ reads
\begin{eqnarray}
r_{123/4}&=&\frac{r_{123}+a_{123} r_{3/4}e^{2i\beta_3d_3}}
 {1-r_{321}r_{3/4}e^{2i\beta_3d_3}},\nonumber\\
 a_{123}&=&t_{123}t_{321}-r_{123}r_{321}=
 \frac{e^{2i\beta_2d_2}-r_{12}r_{32}}
 {1-r_{21}r_{23}e^{2i\beta_2d_2}},\nonumber\\
 t_{123/4}&=&\frac{t_{123}t_{3/4}e^{i\beta_3d_3}}
 {1-r_{321}r_{3/4}e^{2i\beta_3d_3}},
 \label{rA2}
 \end{eqnarray}
 where $r_{123}$, $r_{321}$, $t_{123}$ and $t_{321}$ are given
 by Eq. (\ref{r12-3}).
 In our consideration of the Casimir effect in a planar cavity
 \cite{Tom02} we have used this result to calculate the reflection
 coefficient of a slab in front of a cavity mirror.

As mentioned, grouping the layers is particularly useful when
calculating Fresnel coefficients of periodic media. Consider, for
example, the reflection coefficient of the central segment
$121\ldots 121$ of a Bragg mirror formed by two alternating
quarter-wavelength ($d_i=\lambda/4n_i$) dielectric layers 1 and 2.
Noting that at normal ($k=0$) incidence $e^{2i\beta_id_i}=-1$  and
$a_{121}=-1$, we find from Eq. (\ref{rA2}) (with 3 and
$4\rightarrow 1$) that the normal reflection coefficient $R_N$ of
the segment with $N$ type 2 layers can be calculated using the
algorithm: $R_N=(R_1+R_{N-1})/(1+R_1R_{N-1})$,
%\begin{equation}
%r_N=\frac{r_1+r_{N-1}}{1+r_1r_{N-1}},
%\end{equation}
with $R_0=0$ and $R_1\equiv r_{121}=2r_{12}/(1+r_{12}^2)$ as
input. Of course, grouping the layers in a different way leads to
a different algorithm. Thus, one may show that starting with $R_2$
the following algorithm holds: $R_{4N}=2R_{2N}/(1+R^2_{2N})$,
where in each step the number of type 2 layers is doubled
\cite{ref}.

\begin{figure}[htb]
 \begin{center}
 \resizebox{6cm}{!}{\includegraphics{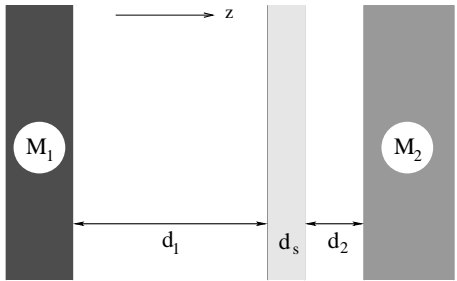}}
 \end{center}
 \caption{\label{cav}A slab (s) in a planar cavity ($n_1=n_2=n_c$)
 schematically. Cavity mirrors are described by their reflection
 coefficients $R_1$ and $R_2$ and the slab by its Fresnel coefficients
 $r \equiv r_{1/2}=r_{2/1}$ and $t\equiv t_{1/2}=t_{2/1}$.}
\end{figure}

Evidently, the above nonstandard recurrence relations are
particularly convenient when the Fresnel coefficients of a stack
are already known (being either calculated separately or measured)
or are to be calculated by some method at a later stage. To
illustrate this, we (re)derive and generalize to media with arbitrary
properties the formula for the Casimir force on a slab in a planar
cavity \cite{Tom02}. Referring to Fig. \ref{cav} for the description
of the system,
the Casimir force (per unit area) acting on the slab is given by
\cite{Tom02}
\begin{equation}
F=T^{(2)}_{zz}-T^{(1)}_{zz},\nonumber
\end{equation}
\begin{equation}
\label{Fj} T^{(j)}_{zz}=\frac{\hbar}{2\pi^2}\int_0^\infty d\xi
\int^\infty_0 dkk\kappa\sum_{p,s} \frac{r_{j-}r_{j+}e^{-2\kappa
d_j}} {1-r_{j-}r_{j+}e^{-2\kappa d_j}}
\end{equation}
with $T^{(j)}_{zz}$ being the relevant component of the
vacuum-field (Minkowski) stress tensor in the cavity region $j$.
Here $\kappa=\sqrt{n^2_c(i\xi)\xi^2/c^2+k^2}$ is the perpendicular
wave vector at the imaginary frequency ($\omega=i\xi$) in the
cavity and $r_{j\pm}(i\xi,k)$ are the reflection coefficients of
the right and left stack of layers bounding the region $j$. We
observe that according to the identity Eq. (\ref{DD}) the tensor
components $T^{(1)}_{zz}$ and $T^{(2)}_{zz}$  are related to each
other \cite{Tom02}.

Noting that $r_{1-(2+)}=R_{1(2)}$ and using the recurrence
relation [cf. Eq. (\ref{rj-k-m})]
\begin{equation}
r_{1+(2-)}=r+\frac{{t}^2R_{2(1)}e^{-2\kappa d_{2(1)}}}
{1-rR_{2(1)}e^{-2\kappa d_{2(1)}}},
\end{equation}
$F$ can be expressed as
\cite{Tom02}
\begin{equation}
\label{F} F=\frac{\hbar}{2\pi^2}\int_0^\infty d\xi \int^\infty_0
dkk\kappa\sum_{p,s}r \frac{R_2e^{-2\kappa d_2}- R_1e^{-2\kappa
d_1}}{N},\nonumber
\end{equation}
\begin{eqnarray}
N&=&1-r(R_1e^{-2\kappa d_1}+R_2e^{-2\kappa d_2})\nonumber\\
&+& ({t}^2-{r}^2)R_1R_2e^{-2\kappa (d_1+d_2)}.
\end{eqnarray}
Since properties of the slab and mirrors are not specified this
result is valid for arbitrary media. For local uniform media
described by the corresponding refraction indexes, Fresnel
coefficients $r=r_{1s2}$ and $t=t_{1s2}$ are given by Eq.
(\ref{r12-3}) and those of the mirrors by Eq. (\ref{sic}). In that
case, of course, the above result agrees with the one obtained
through a conventional way \cite{Elli}. When removing a mirror
(letting, say, $d_2\rightarrow\infty$), the above result gives the
Casimir force between two arbitrary planar objects, as also
obtained recently through a different approach \cite{Esq}.

To summarize, we have emphasized and demonstrated the possibility
of grouping the layers into stacks when calculating Fresnel
coefficients of a multilayered system. This enables one to
consider wave propagation in local as well as in piecewise
nonlocal stratified media on an equal footing. As an example, we
have shown that the formula for the Casimir on a slab in a planar
cavity derived considering local media \cite{Tom02} is also valid
when the objects involved have arbitrary properties.

%\acknowledgments
The author is indebted to I. Brevik for useful interactions and
encouragement. This work was supported by the Ministry of Science,
Education and Sport of the Republic of Croatia under Contract No.
098-1191458-2870.


\begin{thebibliography}{00}
\bibitem{ref}This algorithm was formulated upon the referee's suggestion.
\end{thebibliography}
 \end{document}